# A Deep Learning Approach for Detecting Traffic Accidents from Social Media Data

Zhenhua Zhang
Department of Civil, Structural and Environmental Engineering
University at Buffalo, The State University of New York
Buffalo, NY 14260
Email: zhenhuaz@buffalo.edu

Qing He[1]
Department of Civil, Structural and Environmental Engineering
Department of Industrial and Systems Engineering
University at Buffalo, The State University of New York
Buffalo, NY 14260
Email: qinghe@buffalo.edu

Jing Gao
Department of Computer Science and Engineering
University at Buffalo, The State University of New York
Buffalo, NY 14260
Email: jing@buffalo.edu

Ming Ni
Department of Industrial and Systems Engineering
University at Buffalo, The State University of New York
Buffalo, NY 14260
Email: mingni@buffalo.edu

---

[1] Corresponding author

**Abstract**

This paper employs deep learning in detecting the traffic accident from social media data. First, we thoroughly investigate the 1-year over 3 million tweet contents related to traffic accidents in two metropolitan areas: Northern Virginia and New York City. Our results show that paired tokens can capture the association rules inherent in the accident-related tweets and further increase the accuracy of the traffic accident detection. Second, two deep learning methods: Deep Belief Network (DBN) and Long Short-Term Memory (LSTM) are investigated and implemented in extracted tokens. Results show that DBN can obtain an overall accuracy of 85% with about 44 individual token features and 17 paired token features. The classification results from DBN outperform those of Support Vector Machines (SVMs) and supervised Latent Dirichlet allocation (sLDA). Finally, to validate this study, we compare the accident-related tweets with both the traffic accident log on freeways and traffic data on local roads from 15,000 loop detectors. It is found that nearly 66% of the accident-related tweets can be located by the accident log and more than 80% of them can be tied to nearby abnormal traffic data. Several important issues of using Twitter to detect traffic accidents have been brought up by the comparison including the location and time bias, as well as the characteristics of influential users and hashtags.

**Keywords**: traffic accident detection; tweet; social media; association rules; Deep Learning

**1 Introduction**

Traffic accidents disturb the traffic operations, break down the traffic flow, and cause severe urban problems worldwide. Major traffic accidents can sometimes lead to irreparable damages, injuries, and even fatalities. National Highway Traffic Safety Administration (NHTSA), which publishes yearly reports on traffic safety facts, states that since 1988 more than 5,000,000 car crashes occur in the States each year and about 30% of them result in fatalities and injuries (NHTSA, 2015). After years of research, it has been widely accepted that significant reductions of accident impact can be achieved through effective detection methods and corresponding response strategies. As an essential component of traffic incident management, accurate and fast detection of traffic accidents are critical to modern transportation management (He et al., 2013).

Traditional detector-based methods usually give accurate location and time of the traffic accident and have been proved valid in many applications (Hall et al., 1993; Samant and Adeli, 2000; Sethi et al., 1995). Despite the adaptabilities of previous studies, traditional detection methods with only traffic data still meet certain challenges. First, most of the previous research, which utilized the field data to detect the traffic accidents, build on an implicit assumption that the data is reliable. However, detector failures and communication errors are perennial problems in traffic operations. For example, Illinois Department of Transportation (IDOT) in Chicago reported that around 5 percent of their loops (detectors) are inoperative at any given time (Kell et al., 1990). The problem of malfunctioned sensors could cause even more troubles in accident detection in large regions. Second, the uncertainty nature of traffic patterns and non-recurrent events may undermine the potential of traffic metrics in justifying the traffic accidents. Besides traffic accidents, daily traffic operations may suffer breakdowns by other factors such as parades, road construction, running races, etc. Thus, the metrics including the traffic flow and occupancy inherently perform as an indirect support for traffic accidents instead of direct proof. To address these challenges, there are efforts in applying clustering or classification methodologies such as K-means (Münz et al., 2007) on large data collections to diminish the errors. Our counter-measure lies in extracting "direct report" from tweet users, and the applicability is fully discussed. At the same time, one of the traditional methods is also employed to validate our results in later sections.

In recent years, the accident-related studies have witnessed the power of data crowdsourcing in complementing the traditional methods and finding new knowledge. In this study, we explore the possibility of using Twitter to detect the traffic accidents. Twitter, the microblogging service that received increasing attention in recent years, has been gradually accepted as a direct user-contributed information source in event detection. Twitter establishes an online environment where the content is created, consumed, promoted, distributed, discovered or shared for purposes that are primarily related to communities and social activities, rather than functional task-oriented objectives (Gal-Tzur et al., 2014). Thus, each tweet acts as a data source of "We Media", and it is entirely possible to retrieve the wide-range information from the broad masses of people in a timely manner. Our preliminary examinations also demonstrate the potential of Twitter in delivering the accident-related information.

Table 1 Tweet samples describing the general traffic information, general traffic incident and road accident, respectively

| | |
|---|---|
| General information | *"I am waiting at the silver line, exciting"* |
| | *"Always hate the signals ahead of the hip-hop, making me sick"* |
| General incident | *"standstill for 1 hour, there must be accidents in front"* |
| | *"this is typical NOVA traffic, what a bad day"* |
| Traffic accident | *"major accident next to the sunoco near the parkway a car got flipped over"* |
| | *"the worst car accident possible just happened in front of me"* |

As we can see from Table 1, the Twitter information is both noisy and unstructured. An effective text mining method is necessary to extract the useful accident-related information from tweets. In this study, we employ and compare two deep learning methods: Deep Neural Network (DNN) and Long Short-Term Memory (LSTM), in training and classifying the accident-related tweets. Unlike classifiers such as logistic regression, Support Vector Machines (SVMs) or Artificial Neural Network (ANN) with a single hidden layer, deep learning does not seek direct functional relationships between the input features and the output classification results. Instead, it is a set of machine learning algorithms that attempt to learn in multiple levels, corresponding to different levels of abstraction (Deng and Yu, 2014). The training process of DNN is divided into multiple layers, and the output result is expressed as a composition of layers, where the higher level features are the composition of lower-level features, giving the potential of modeling complex data with fewer units than a similarly performing shallow network (Bengio, 2009). Our efforts can be detailed by three major contributions: First, we propose a systematic feature selection process in extracting both the individual and paired token features from social media. We unveil the language customs of the tweet users in describing the traffic accident detection. Second, we validate the effectiveness of the deep learning approach in classifying the social media data. The results show that deep learning outperforms other prevailing data mining methods. Third, the advantages and disadvantages of tweets in accident detection are verified and fully discussed by comparing tweets with both accident log from state Department of Transportation and traffic data from thousands of loop detectors.

The rest of the paper is organized as follows: Section 2 reviews the current studies in social media applications in transportation and the deep learning in language modeling. Section 3 introduces tweet data preprocessing for accident detection. Section 4 details the process of feature selection from tweet contexts including both the individual token features and paired token features. Section 5 introduces the two deep learning method: DBN and LSTM; and their performances in classifying the tweets as compared with SVMs and sLDA. Section 6 validates the accident-related tweets by the traffic accident log and loop-detector data. Section 7 concludes the paper with a few empirical findings and generalizations together with some thoughtful discussions.

## 2 Literature review

### 2.1 Review of social media in traffic-related studies

The newly emerged data source, social media data, has proved its capability in recent traffic studies including activity pattern identification (Hasan and Ukkusuri, 2014), special traffic-related events (Ni et al., 2014; Shirky, 2011), traffic flow prediction (Cottrill et al., 2017; Lin et al., 2015; Ni et al., 2017), transport information management (Cottrill et al., 2017), travel mode detection (Maghrebi et al., 2016), destination or route choice (Huang et al., 2017), etc. According to Rashidi et al. (2017), as social media data encompasses information that is revealed by users in realistic situations, such data is free from sampling, surveying or laboratory biases. The location effectiveness and timeliness features of Twitter can be proved in a recent accident detection study that uses the GPS-enabled smartphones (White et al., 2011) and travel behavior study which has been validated by the household travel survey (Zhang et al., 2017).

Studies related to incident detection are good at leveraging the location and time information from tweets to deliver their research goals: Schulz et al. (2013) used microblogs to detect the small-scale incidents. Gal-Tzur et al. (2014) conducted a corridor study on the correlation between tweet and traffic jam. D'Andrea et al. (2015) compared accuracies and precisions of different regression models including Naïve-Bayes, Support Vector Machine, Artificial Neural Network, Decision Tree in detecting traffic incidents from Twitter stream. Gu et al. (2016) combined the data sources from Twitter, incident records, HERE, etc. and employed the Naïve-Bayes classification to detect five major incident types.

Some of the above studies also mentioned the challenges to be addressed for tweets in traffic accident detection. First, as compared to events that arouse enormous public attention such as sporting games, extreme weathers or traditional festivals, the influence of traffic accidents are comparably a "midget" (Lin et al., 2015; Zhang et al., 2016c). From our observations, accident-related tweets are thus in small quantity. What's more, most of them are confined to a small area and limited to a relatively short time interval, and some researchers call them small-scale events (Schulz et al., 2013). It is worth exploring the effectiveness and limitations of tweets in detecting small-scaled events, especially the features of timeliness, accuracy, etc. Second, another challenge in tweets lies in its inherent complexity and unstructured nature of data: language ambiguity (Chen et al., 2014). The common methods for detecting the traffic-related events include Support Vector Machine (D'Andrea et al., 2015; Schulz et al., 2013), natural language processing (Li et al., 2012; Wanichayapong et al., 2011), etc. which explore the semantic features in the keywords. However, as the context of a tweet is limited to 140 words and the tweet contents try to be concise, keyword detection is sometimes not sufficient for accurate automatic language processing. For example, "internet traffic is slow" and "internet shows traffic is slow" may deliver totally different information. Third, also due to the word limitation, some tweet contents that do not give enough descriptions of the incident types, even if some of the incidents may come from their suppositions. Previous works had been successful using keywords in Twitter to analyze the sentiment of rider dissatisfaction along the designated routes (metro lines) (Collins et al., 2013). In comparison, natural language for reporting traffic incidents can be more diverse and difficult to capture. These three challenges are the major interests of this paper, and they will be statistically proved and fully discussed in later sections.

**2.2 Review of machine learning in the language modeling**

The machine learning methods have thrived in the applications of language and text modeling in recent years, which can potentially counter the challenges in processing and classifying the tweets. In most of the studies, language modeling can be taken as a kind of information extraction from the text messages, which is the process of converting the unstructured text information into a structured database and solving it as a supervised or unsupervised learning task. The limited word features can be utilized for specific research. For instance, Tong and Koller (2001) proposed a new algorithm for performing active learning with Support Vector Machine for text classifying; Campbell etc. (2006) also employed the Support Vector regression for speaker and language recognition; There are also algorithms based on Naive Bayes classifier for text classification or language modeling. In an early study by McCallum and Nigam (1998), the comparison between two models make the "Naive Bayes assumption" show that the multinomial model is found to be almost uniformly better than the multi-variate Bernoulli model. From the view of sLDA, a tweet post can also be disintegrated into a bag of topics; and the proportions of those topics can be approximated by a distribution (e.g. Dirichlet distribution) and even be inferred from word distribution in each topic.

Besides, the newly emerging methods: deep learning have attracted increasing attention and have been proved to be superior in some transportation studies. For instance, deep learning architecture has been proven better than the artificial neural network in traffic flow prediction (Polson and Sokolov, 2017). Its advantages in supervised learning lie in setting additional layers between the input and output, which replaces handcrafted features with efficient algorithms and hierarchical feature extraction (Song and Lee, 2013).

Attempts of employing deep learning method in text classification thrive in many branches including Deep Belief Network (DBN), Recurrent Neural Network (RNN), Convolutional Network (Zhang et al., 2015), etc. DBN is the composition of simple network such as restricted Boltzmann machines (RBMs) (Hinton, 2009), etc. RNN can be taken as multiple copies of the same network and are able to pass information in sequence from the previous inputs to the present task. Thus, it is a powerful model for sequential data and proves valid in long speech recognition (Graves et al., 2013). If we employ RNN for a supervised learning task in language modeling, the process can be described as using a sequence of word features to predict the manual labels such as topics, sentiment (Agarwal et al., 2011), etc. One special form of RNN: Long Short-Term Memory Network (LSTM) (Hochreiter and Schmidhuber, 1997) moves one step further which is capable of learning long-term dependencies between words within the context. The LSTM unit in the network can remember the inputs for either long or short durations. The input information from lower layers is neither converted nor eliminated because there is no conversion from lower layers to upper layers. These unique features are valued in the applications such as speech tagging (Wöllmer et al., 2010) or handwriting recognition (Graves et al., 2009). Applications using RNN or LSTM for classification is an attractive choice for sequence labeling (Graves, 2012), which can finish a variety of tasks in topic modeling.

Given that DNN models work well in language modeling, we expect them to deliver promising results in classifying the accident-related tweets. This is mainly because DNN consists of multiple layers or stages of nonlinear information processing which captures the inter-feature correlation. Besides, it can represent features successively by higher, more abstract layers. The deep learning methods are expected to be more effective in classifying the accident-related tweets than other methods. Two networks are tested in this studies: DBN and LSTM. In addition, their effectiveness in classifying short tweet contexts will be fully discussed.

## 3 Data description and preprocessing

### 3.1 Raw data and study area

We have two study areas. The first metropolitan area is Northern Virginia (NOVA). With 2.8 million residents (about a third of the state), NOVA is the most populous region of Virginia and the Washington D.C. Metropolitan Area. The road network is a 2500 km$^2$ area with more than 1,200 signalized intersections and has long been known for its heavy traffic (Zhang et al., 2016a). The second study area is the New York Metropolitan Area (NYC), in which the population amounts up to 23.7 million (U.S.Census.Bureau, 2016). The area can be even larger up to 3000 km$^2$.

The tweet data were collected through Twitter Streaming API with geo-location filter. Filtering by the coordinates, we can set bounding boxes in our study areas obtaining more than 584,000 geo-tagged tweets in NOVA and 2,420,000 in NYC in a full year from January 2014 to December 2014. Each tweet posts are coupled with specific date, time and location information. The location information is the paired latitude and longitude where the tweets are posted.

### 3.2 Tweet data preprocessing

In this study, we only study the tweets which have explicit indications of traffic accidents and do not include the traffic congestion, construction works, etc. because the latter may not necessarily indicate traffic accidents. The types of the traffic accidents in this paper can be summarized into 3 categories: "collision", "disabled vehicle" and "vehicle on fire".

The raw tweet data need to be preprocessed to constitute the database that can be used for further analysis. The first step is to extract the candidate tweets that possibly describe the on-site traffic accidents. Usually, these candidate tweets should contain one or more keywords such as "accident" or "crash" that are accident-related. However, there has been no consensus on such a vocabulary of the accident-related words. Thus, we turn to the traditional news media and collect about 100 articles of news that discuss the traffic accident. In all these articles, we select the words that appear the most frequently. The frequency of a word is the times that a specific word appears in these articles. Except for the common words such as "I", "is", etc. and those that reflect specific geographic and event features, we found that most of the articles contain a common list of words with a high frequency as shown in Table 2.

Table 2 Accident-related words

| police, accident, traffic, crash, road, car, vehicle, highway, driver, county, injured, pm, state, injuries, scene, hospital, according, people, died, near, patrol, morning, happened, dead, taken, just, driving, department, involved, vehicles, south, passenger, hit, truck, north, monday, left, lanes, lane, killed, struck, southbound, area, closed, investigation |
| --- |

The second step is to extract the candidate tweets based on these accident-related words. By applying a filter based on keywords in Table 2, we can obtain a large quantity of potential tweets. On these tweets, we looped the following procedures to filter out the non-accident-related tweets and obtain the related words:

- Randomly select tweets from the filtered tweets.

- Manually label them whether they are accident-related.
- Extract the most frequent words in accident-related tweets.
- Filter the tweets based on the frequent words.

As compared to traditional media, social media blogs are short, brief with few editorial review. Some of the words may have grammatical errors. Also, some tweets posted by the influential users do not provide valid location information. Thus, to ensure both the accuracy and sample size, certain rules are further implemented:

- Include the words that are relevant to accidents but apparently misspelled or personally modified including "acident", "incdent", etc.
- Include other variations of accident-related words such as the word pairs that have a hyphen in word pair such as roll-over, etc.
- Exclude the tweets posted by influential users including web media, transportation authorities or Department of Transportation listed in Table 3. Note that some public tweets may mention the names of influential users or hashtags and should not be deleted. Also, there are some tweets posted by the personal Twitter account of the reporters working for the authorities. They may still give valid location and time information and should not be deleted.

Table 3 Names of hashtags and influential users found in tweets

| | |
|---|---|
| Influential user | nbcnews, fox5newsdc, wtoptraffic, washingtonpost, wtop, nbcwashington, wtoptraff, abc2news, metlife, traffic_nyc, wcbs880traffic, skywayrehab, totaltrafficnyc |
| Hashtag | onlyindc, dctraffic, vatraffic, mdtraffic, vatraff, nyctraffic, nyc, njtraffic |

Finally, we obtained more than 900 accident-related tweets in our study areas. To generate a balanced dataset, we randomly select non-accident-related tweets which are twice the size of accident-related tweets and combine them into the structured tweet database.

**3.3 Structured database construction**

Each tweet post can be decomposed of words, characters, numbers or even Latinized symbols that are collectively called "tokens". We can find more than 20000 tokens from all tweets. Some of them will be selected as the features after necessary filtering and stemming. The procedures can be illustrated in Figure 1.

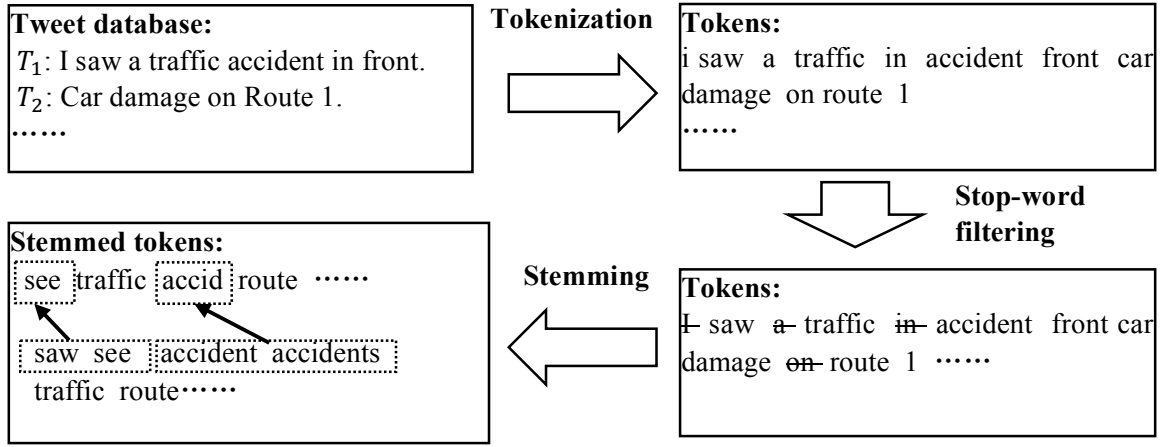

Figure 1 Steps of token filtering and stemming

First, the punctuation marks convey almost no meanings and should be discarded, and all other words should be converted into lower case. Meanwhile, some of the words or characters that have no apparent linguistic meanings or significant event indications should be filtered out before the processing. The stop-word list we used refers to Ranks-NL (2015). Some of the tweets also contain the names of hashtags or influential users shown in Table 3. These names are usually mentioned by "#" and "@". In this paper, we did not include them as the features. This is because these hashtags relate closely to certain areas and are more likely to become features than other words; the models built on the hashtags will then be less effective when applied to other areas. Second, some of the words have different writing expressions but convey almost the same meanings such as "accidents" and "accident". The token stemming is necessary to reduce these inflected (or sometimes derived) words to their word stem, base or root form. In this study, we employ the Porter stemming algorithm (Porter, 1980) for the token stemming.

After token filtering and stemming, each tweet $T_i$ can be disintegrated and summarized by a set of stemmed tokens. Of all the tweets, the stemmed tokens are labeled as $[t_1, t_2, ... ... t_j]$. The stemmed tokens are the features for each tweet $T_i$ and each tweet has different token features. If the tweet contains a stemmed token, the corresponding token features are labeled as 1, otherwise 0. After this, the unstructured tweet sentences $T$ are converted into a structured binary database $D^S$ for further feature analysis and text classification.

**4 Feature Selection**

This section selects the features from the structured tweet database. The main idea of Feature Selection (FS) is to select a subset of features from the original documents. FS is performed by keeping the words with the highest score according to a predetermined measure of the importance of the word (Korde and Mahender, 2012).

**4.1 Individual token features**

The benchmark we choose for individual token features is phi coefficient (Cramér, 1999), which can measure the association between manual label and tokens. The coefficient (usually denoted as $\phi$) between two variables $x$ and $y$ is calculated as:

$$\phi = \frac{n_{11}n_{00} - n_{10}n_{01}}{\sqrt{n_{1*}n_{0*}n_{*0}n_{*1}}} \quad (1)$$

Where all notations are defined in the following table:

|       | $y = 1$ | $y = 0$ | Total |
|-------|---------|---------|-------|
| $x = 1$ | $n_{11}$ | $n_{10}$ | $n_{1*}$ |
| $x = 0$ | $n_{01}$ | $n_{00}$ | $n_{0*}$ |
| Total | $n_{*1}$ | $n_{*0}$ | $n$ |

Those tokens whose $|\phi|$ is higher than 0.1 are selected. Following this rule, 27 tokens are selected and some of them are shown in Figure 2.

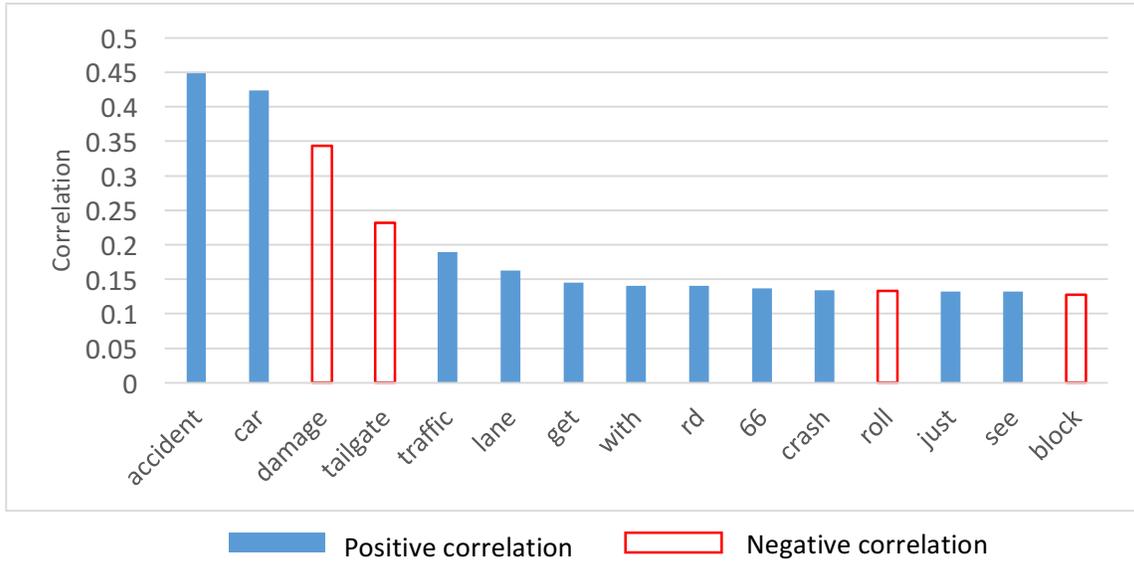

Figure 2 Correlations between the manual label and the individual stemmed tokens. To making it easy to read, we write basic form of the word instead of stemmed token

From Figure 2, the stemmed tokens may be different from their original words in which "accident" converts to "accid". For display purposes, we write the basic form of the words. Some of the tokens may be accounted by the geographic uniqueness such as "66", "95", and "495", indicating the route number where traffic accidents occur. Note that these tokens took a small portion and their correlation values are not as high as the hashtags, thus the models will not lose generalization. Some may be directly topic-related words including "traffic", "car", "accident", etc. Other words such as "damage" and "tailgate" are too general in our daily life and thus provide negative indicators in describing the traffic accident.

**4.2 Paired token features**

Features from individual tokens are sometimes not sufficient because these emphasize solely the correlations between label and tokens and may overlook the associations within the tokens. Sometimes, such associations can have much more significant indications than single ones. For example, in a tweet post, the occurrence of word "car" conditioned by "accident" may increase the accident-related probability. Conversely, the occurrence of token "car" conditioned by "maintenance" or "repair" may decrease the probability. In this section, we select the paired token features by studying the association rules between the manual label and the stemmed tokens in the binary database $D^S$. The association rules can be unveiled by the Apriori algorithm (Agrawal and Srikant, 1994; Hahsler et al., 2007). Apriori algorithm finds the regularities in large-scale binary data by two major probabilities: support and confidence.

Given a stemmed token $t_j$ in all stemmed tokens $[t_1, t_2, \ldots, t_j, \ldots, t_N]$, support of $t_j$ is the proportion of tweets which contain $t_j$ in the database.

$$supp(t_j) = \frac{sizeof(\{T_i, t_j \subseteq T_i\})}{sizeof(\{T_i\})} \tag{2}$$

Where $t_j$ is the $j$th token; $T_i$ is the $i$th tweet. Setting a threshold of $supp(t_j)$, we can filter out a limited number of qualified $t_j$. Similar to the support of each individual token, we can even calculate the support of paired tokens $supp(t_{j_1} \cap t_{j_2} \cap \ldots \ldots \cap t_{j_m})$:

$$supp(t_{j_1} \cap t_{j_2} \cap \ldots \ldots \cap t_{j_m}) = \frac{sizeof(\{T_i, t_{j_1} \cap t_{j_2} \cap \ldots \ldots \cap t_{j_m} \subseteq T_i\})}{sizeof(\{T_i\})} \tag{3}$$

Where $j_1 \neq j_2 \neq \cdots \neq j_m$. One can see that support deals mainly with the frequencies of one or more tokens. As the tweet database are filtered according to several different keywords, the word combinations of accident-related tweets may also be quite different. Thus, support of paired tokens can possibly capture different concurrent tokens that can possibly be used as the features in the model. But not all of them may be qualified as the features in the model. Besides support, the association rule between manual label and the paired tokens can be further revealed by confidence, defined as:

$$conf(L_i \Rightarrow t_{j_1} \cap t_{j_2} \cap \ldots \ldots \cap t_{j_m}) = \frac{supp(L_i \cap t_{j_1} \cap t_{j_2} \cap \ldots \ldots \cap t_{j_m})}{supp(t_{j_1} \cap t_{j_2} \cap \ldots \ldots \cap t_{j_m})} \tag{4}$$

Where $L_i$ represents the accident label. In the confidence calculation, we focus more on paired tokens that are related to traffic accident. The number of individual tokens in a paired token is always more than 1 and the maximum can theoretically be equal to the number of tokens in a tweet post. Also, if one increases the size of the paired tokens, the computational time will dramatically increase bringing almost no benefit.

In most of the previous studies, setting support and confidence is sometimes mandatory. The setting of support can be a small value that can include as many as paired tokens for Feature Selection. The setting of confidence, as compared, usually influences the results significantly and different values should be further investigated in the classification for the impacts. We conduct an empirical study to see how the token features can reveal the language of customs of tweets in describing traffic accidents. When support is equal to 0.01 and confidence is equal to 0.1, we can find 38 token pairs listed in Table 4. Table 4 shows that most of the association rules contain the accident-related tokens as discussed in Section 4.1. The token combinations should conform to the language customs of tweet users: some of them are the combinations between the accident-related tokens; others have the adverb "just" which is a typical oral expression.

Table 4 Paired tokens selected by the Apriori algorithm

| car*     | with | damage | do    | accident | just     |
|----------|------|--------|-------|----------|----------|
| tailgate | game | car    | crash | car      | accident |

| accident | lane | car | get | i | car | crash |
| accident | traffic | tailgate | i | just | car | get |
| i | roll | i | just | just | accident | get |
| just | get | car | i | i | car | get |
| accident | get | tailgate | stadium | accident | car | get |
| car | just | accident | see | i | accident | get |
| damage | i | accident | drive | i | just | car |
| accident | i | i | do | just | accident | car |
| accident | block | i | crash | i | just | accident |
| car | see | i | get | i | accident | car |
| car | drive | damage | just | | | |

\* To make it easy to read, we write the basic form of the word instead of the stemmed token.

Table 5 shows that by changing the confidence, the number of paired tokens may also change. In the meanwhile, the number of individual tokens that are correlated remains almost unchanged when confidence is higher than 0.6.

Table 5 Paired token statistics with different confidence values

| Confidence | Number of paired token features | Number of individual tokens in the features |
| --- | --- | --- |
| 0.1 | 38 | 18 |
| 0.2 | 38 | 18 |
| 0.3 | 38 | 18 |
| 0.4 | 38 | 18 |
| 0.5 | 38 | 18 |
| 0.6 | 30 | 17 |
| 0.7 | 23 | 17 |
| 0.8 | 17 | 16 |
| 0.9 | 9 | 10 |
| 1 | 3 | 4 |

Same as individual tokens, the paired token features in the database are equal to 1 if the tweet contains the corresponding paired tokens and 0 otherwise. We will perform the analysis by incorporating paired token features into the regression model.

**5 Classification by Deep Learning**

**5.1 Deep Belief Network (DBN) and Long Short-Term Memory (LSTM)**

The first deep learning method to be implemented is DBN, which is one of the simplest form of deep learning Networks. It consists of densely-connected layers, and each layer has a few neurons that represent the activation function. There exist links between neurons from different levels of layers while there is no link between neurons in the same level of layers. Connections between neurons in the same layer may not be practical and have scalability issues. Thus, the DBN in this paper is also known as Restricted Boltzmann Machine (RBM). The neural functions and basic structure are shown in Figure 3.

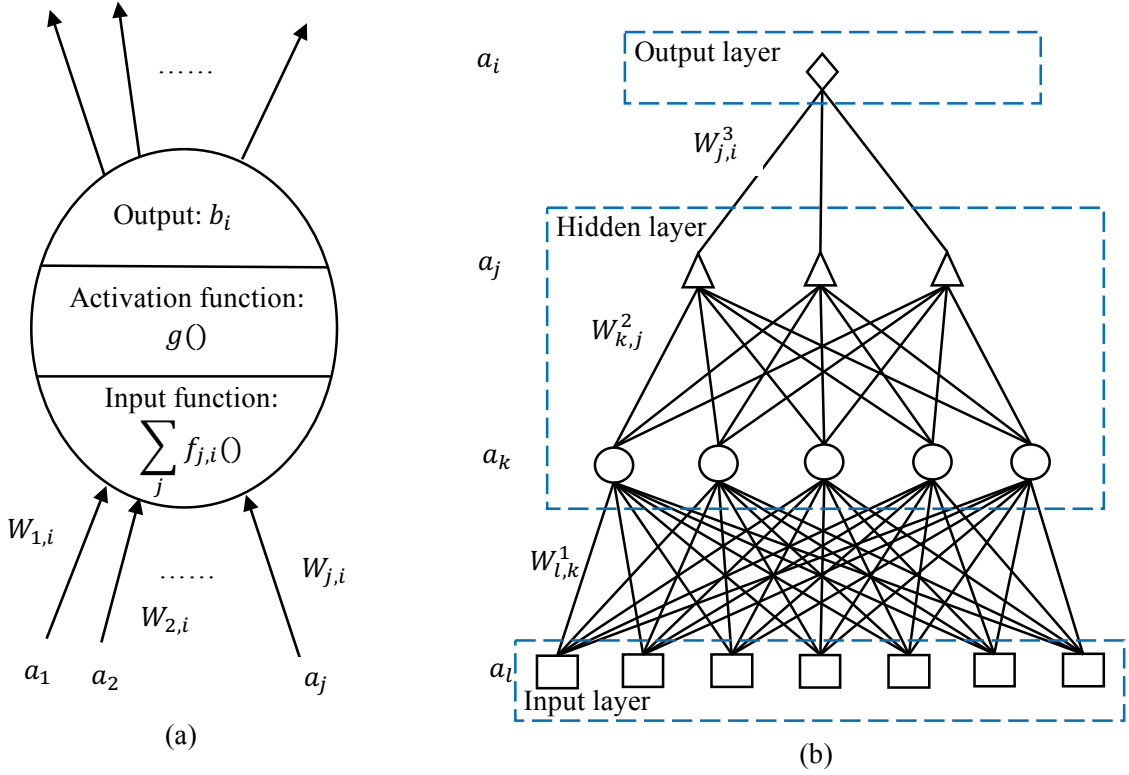

Figure 3 (a) Example of a single neuron. $W_{j,i}$ denotes the transition matrix. (b) Structures of 4-layer neural networks.

The number of categories that a neuron can output is equal to the number of neurons in the upper level. The relationship between the input and output can be written as Equation (5):

$$b_i = g\left(f_{j,i}(W_{j,i}, a_j)\right) = g\left(\sum_j W_{j,i} a_j\right) \tag{5}$$

Where $a_j$ is a one-hot vector for the $j$th input token feature; $b_i$ is the output; $W_{j,i}$ is the conversion parameter matrix to be estimated; $g()$ is the activation function and can be changed in different levels of neuron. The $g()$ can be taken as a series of functions that convert a $i$th dimensional vector (same as $b_i$ in this study) of arbitrary real values into a $i$th dimensional vectors of values in the range between 0 and 1. For instance, the Softmax Function employed in this study can be written in Equation (6):

$$g(f_{j,i}) = \frac{e^{f_{j,i}}}{\sum_i e^{f_{j,i}}} \tag{6}$$

In a 2-layer neural network, the input is the token features while the output is the manual label. The 2-layer neural network resembles the Artificial Neural Network, similar to that of the Support Vector Machines or logistic regression in which the output value is the direct computation of the input features. In a multi-layer neural network, however, the input features are first converted into hidden features as shown in Figure 3(b) and then the hidden layers finally calculate the corresponding output. For the output layer, with an initial estimate of $W_{j,i}$, one can calculate the square error between the true label and estimated label as shown in Equation (7). Thus, the regression problem can be converted into an optimization problem in which can find the best $W_{j,i}$ to minimize the square error $\delta^2$, or diminish the changes of $\delta^2$ until $\nabla(\delta^2)$ smaller than a threshold value by a gradient method or Newton-Raphson method (Ypma, 1995).

$$\delta^2 = \frac{1}{2}\sum_i (y_i - b_i)^2 \tag{7}$$

The gradient method is an iterative approach that in each cycle finds a descent direction and update the $W_{j,i}$ by a step size. The descent direction can be calculated as:

$$\frac{\partial(\delta^2)}{\partial W_{j,i}} = \sum_i (y_i - g_i(\sum_j W_{j,i} a_j)) \cdot \frac{\partial}{\partial W_{j,i}} (y_i - g_i(\sum_j W_{j,i} a_j))$$

$$\frac{\partial(\delta^2)}{\partial W_{j,i}} = \sum_i -e_i g_i' a_j \tag{8}$$

Where $e_i = y_i - g(\sum_j W_{j,i} a_j)$ is the difference between the predicted label and true label; The $W_{j,i}$ can be updated according to the perceptron learning rule (Freund and Schapire, 1999):

$$W_{j,i}^{t+1} = W_{j,i}^t + \alpha \frac{\partial(\delta^2)}{\partial W_{j,i}} = W_{j,i}^t + \alpha \sum_i e_i g_i' a_j \tag{9}$$

Where $\alpha$ is a scale parameter to be decided and $t$ indicates the iteration cycle. $W$ between other layers can be updated in the same way. For the hidden layer, we can update the corresponding $W_{l,k}$ or $W_{k,j}$ by the error from the output layer. The algorithm employed is called back-propagation. The process of back-propagation can be generalized as follows: when the features are placed in the input layer, the effects of the input features are propagated forward through the layer structure, layer by layer until reaching the output layer. By comparing with the true label, using the error function in Equation (7), the error values are then propagated backwards, updating the conversion matrix as shown in Equation (10).

$$\begin{aligned}
\frac{\partial(\delta^2)}{\partial W_{k,j}} &= \sum_i -(y_i - g_i(\sum_j W_{j,i} a_j)) \cdot \frac{\partial}{\partial W_{k,j}} (g_i(\sum_j W_{j,i} a_j)) \\
&= \sum_i -e_i g_i' \frac{\partial}{\partial W_{k,j}} (\sum_j W_{j,i} a_j) \\
&= \sum_i -e_i g_i' W_{j,i} \frac{\partial a_j}{\partial W_{k,j}} \\
&= \sum_i -e_i g_i' W_{j,i} g_j' \frac{\partial g_j}{\partial W_{k,j}} \\
&= \sum_i -e_i g_i' W_{j,i} g_j' \frac{\partial}{\partial W_{k,j}} (\sum_k W_{k,j} a_k) \\
&= \sum_i -e_i g_i' W_{j,i} g_j' a_k
\end{aligned} \tag{10}$$

The algorithm can be generalized as follows:

---

Algorithm of Deep Neural Network

---

Input: Token features: $a_l$;

    Manual labeled data: $y_i$.

Output: Predicted data: $a_i$;

    Transition matrix: $W_{l,k}$, $W_{k,j}$ and $W_{j,i}$

Set iteration $t=1$ and initial guess of $W_{l,k}$, $W_{k,j}$ and $W_{j,i}$

Repeat

    Implement $\frac{\partial(\delta^2)}{\partial w_{j,i}}, \frac{\partial(\delta^2)}{\partial w_{k,j}}$ and $\frac{\partial(\delta^2)}{\partial w_{l,k}}$ as instructed in Equation (8) and (10);

    Choose $\alpha$ and update $W_{l,k}$, $W_{k,j}$ and $W_{j,i}$ as instructed in Equation (9);

Until $\delta^2 \leq threshold$

---

One can see that the computation workload by a greedy learning algorithm (Hinton et al., 2006) can be quite heavy given so many neurons and activation functions. By setting a proper number of neurons in a layer, one can obtain a good fit with a reasonable computing time. Proper reduction of dimensionality (Hinton et al., 2006) is essential to accelerate the process. The computing time and the setting of neurons will be discussed in this paper. Besides the computing time, we further implement 5-fold cross validation to overcome the over-fitting problem (Geisser, 1993) in the process of model training. Cross-validation can give insights in how the model will generalize to an independent dataset. Directed by this method, the dataset is randomly partitioned into 5 folds. The classification model is trained on 4 folds, and the remaining fold is used for testing the trained model. This procedure is repeated 5 times and each fold is used exactly once as a test data. We can finally obtain an overall estimation by averaging 5 test results.

The second deep learning network, Long Short-Term Memory Network (LSTM), is the newly emerged deep learning method applied to language modeling. In previous works such as (Graves, 2012), the supervised learning using LSTM is taken as a form of sequence labeling. The general form of LSTM is shown in Figure 4(a) in which the both network structure and input originates from the Recurrent Neural Network (RNN). Unlike DBN, RNN creates an internal state which retains the sequential input information, and LSTM is a kind of modified RNN which can solve the long-term dependency problem in language modeling. In LSTM, each token feature is put into the network by a time step $t$. At each time step, $x_t$ is put into a chunk of neural network $a_t$ and produces a temporary output $h_t$. The information $x_t$ carries will be retained in $a_t$ and transferred to the network chunk in the next step $a_{t+1}$.

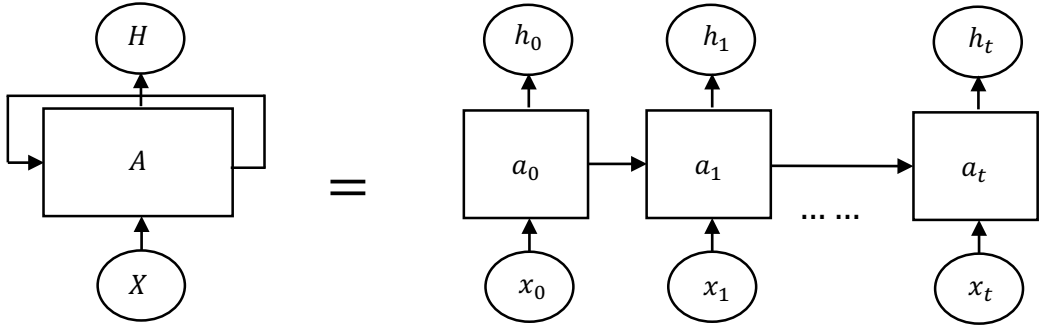

(a)

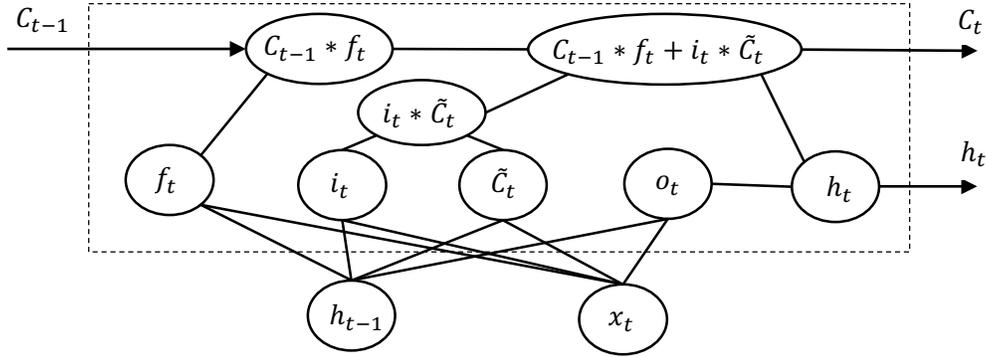

(b)

Figure 4(a) The general form of Recurrent Neural Network; (b) The network structure of Long Short-Term Memory at step $t$

There are different forms of LSTM in which $a_t$ is structured in different ways and different kinds of LSTM serve different research purposes. However, there are studies doing a comparison between some variants and the results show little difference (Greff et al., 2016). Our study adopts one of the basic form of LSTM first proposed by Hochreiter and Schmidhuber (1997). In $a_t$, the input at step $t$ includes the a token feature $x_t$ and classification result $h_{t-1}$ at step $t-1$. There is also a cell $C_{t-1}$ which can retain the information in the previous inputs. The network structure functions as shown in Figure 4(b). In each step, $C_t$ can be updated by Equation (11).

$$C_t = f_t * C_{t-1} + i_t * \tilde{C}_t$$

*where*

$$f_t = g(W^f \cdot [h_{t-1}, x_t] + b^f)$$
$$i_t = g(W^i \cdot [h_{t-1}, x_t] + b^i)$$
$$\tilde{C}_t = tanh(W^C \cdot [h_{t-1}, x_t] + b^C) \qquad (11)$$

In Equation (11), [,] combines the two vector by column; $g()$ is the activation function similar to Equation (6): $g(x) = \frac{1}{1+e^x}$. The $h_t$ can be updated by Equation (12).

$$h_t = o_t * \tanh(C_t)$$

where

$$o_t = g(W^o \cdot [h_{t-1}, x_t] + b^o) \tag{12}$$

As LSTM is designed to classify the sequential input, the order of the token feature $x_t$ should be accordant with those in the tweets. During the model training, back-propagation algorithm can estimate $W^f, W^i, W^C, W^o$ and their corresponding $b$ by calculating their partial derivatives which is a similar process as DBN. 5-fold cross validation is also necessary and the functional API: Keras is employed to define the complex LSTM models and fine-tuning with different input features.

## 5.2 Classification results and comparisons

### 5.2.1 Classification results of DBN

To evaluate the achieved results, we employ statistical metrics: accuracy and precision.

$$Accuracy = \frac{TP + TN}{TP + FP + FN + TN} \tag{13}$$

$$Precision = \begin{cases} \dfrac{TP}{TP + FP} & for\ accident-related \\ \dfrac{TN}{TN + FN} & for\ non-accident-related \end{cases} \tag{14}$$

where TP, TN, FP, and FN are defined as follows,

|  | $Prediction = 1$ | $Prediction = 0$ |
|---|---|---|
| $Groundtruth = 1$ | $TP$ | $FP$ |
| $Groundtruth = 0$ | $FN$ | $TN$ |

In Equation (14), precision is calculated separately for accident and non-accident tweets.

By setting the confidence to be 0.8 for Feature Selection of paired tokens, there will be 17 paired token features and totally 16 individual tokens in the paired token features. By combining these paired token features and the individual token features, we can finally obtain good regression results as shown in Figure 5. As discussed in Section 4.1, The number of the individual token features in the regression model changes with the correlation coefficient $\phi$. When we set $\phi$ to be 0.2, there are only 4 qualified individual token features and the accuracy can be around 0.8. Higher $\phi$ may result into a simpler model but relatively less accuracy; while lower $\phi$ improves the performance but may cause overfitting. Thus, one may seek a balanced model in the future applications.

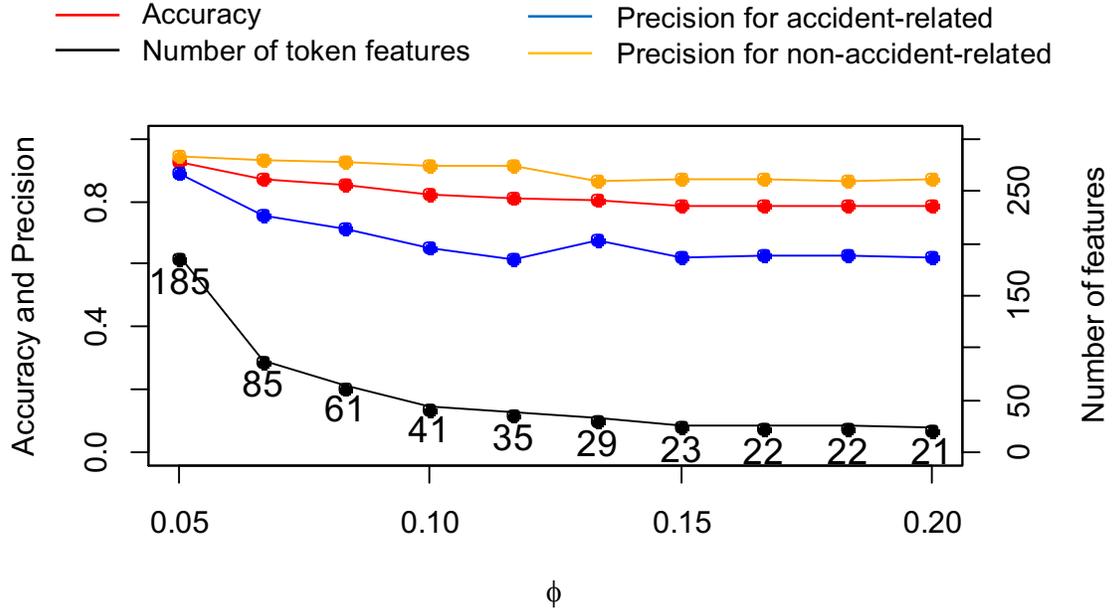

Figure 5 Regression results of DBN with selected individual and paired tokens under different thresholds of correlation coefficient $\phi$

Incorporating the paired token features in classifying the tweets has certain advantages. Table 6 shows an increase in accuracy with paired token features during different stages of $\phi$.

Table 6 Accuracy results with both token features and individual token features only

| $\phi$ | With both token features | With only individual token features | Difference |
|---|---|---|---|
| 0.05 | 0.925 | 0.922 | 0.32% |
| 0.067 | 0.872 | 0.871 | 0.15% |
| 0.083 | 0.854 | 0.849 | 0.65% |
| 0.1 | 0.825 | 0.826 | -0.10% |
| 0.117 | 0.813 | 0.805 | 0.95% |
| 0.133 | 0.803 | 0.795 | 1.07% |
| 0.15 | 0.787 | 0.779 | 1.09% |
| 0.167 | 0.788 | 0.766 | 2.89% |
| 0.183 | 0.787 | 0.766 | 2.78% |
| 0.2 | 0.787 | 0.751 | 4.88% |

The parameter setting may influence the accuracy and precision but not to a large extent. Figure 6 shows by increasing the number of neurons in the second and third layers, the regression results do not change greatly while computation time increases a lot. The computing time matters since it will be a major concern for real-time applications.

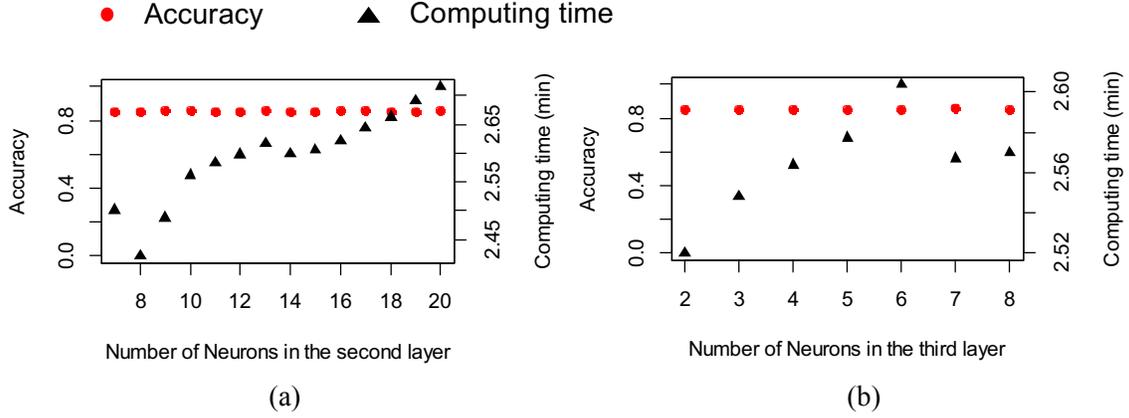

Figure 6 Regression results and computing times by (a) different number of neurons in the second layer when the third layer has 5 neurons; (b) different number of neurons in the third layer when the second layer has 10 neurons. $\phi$ is equal to 0.117 in both cases.

**5.2.2 Comparing DBN with LSTM, ANN, SVMs and sLDA**

This section compares the classification results of the DBN with LSTM as well as a supervised learning method: Artificial Neural Network (ANN), Support Vector Machine (SVMs) (Karatzoglou et al., 2005) and a topic modeling method: supervised Latent Dirichlet allocation (sLDA) (Mochihashi, 2009). ANN employed in this comparison is a feed-forward neural network with a single hidden layer (Venables and Ripley, 2013); the number of nodes in the hidden layer is equal to 5 which is the same as that in DBN. SVMs is a supervised learning model and can employ different kernel functions to keep the computational load reasonable. In this comparison, we employ the linear kernel to train and predict the models. 5-fold cross-validation is also employed in the process of model training to avoid overfitting.

The supervised Latent Dirichlet Association assumes that a topic is a probability distribution over a group of words (tokens) which describe a semantic theme and the features of a document can be divided into several different topics instead of different words (tokens). Thus, sLDA is capable of reducing the dimensionality of the words. As compared most of the topic models including Latent Dirichlet Allocation (LDA) which are unsupervised, sLDA can infer latent topics of the response on the basis of a manual label. The advantages of sLDA have been proved in several studies. According to Mcauliffe and Blei (2008), each tweet post and label are processed from the following generative process:

- Draw topic proportions $\theta|\alpha \sim Dir(\alpha)$;
- For each word
    - Draw topic assignment $z_n|\theta \sim Mult(\theta)$;
    - Draw topic assignment $w_n|z_n, \beta_{1:K} \sim Mult(\beta_{z_n})$;
- Draw response variable $y|z_{1:N}, \eta, \sigma^2 \sim Mult(\eta^T \bar{z}, \sigma^2)$.

Where $Dir(\alpha)$ is the Dirichlet distribution; $Mult(\theta)$ is the multinomial distribution; $z_n$ is the topic of the word $w_n$ (token); $\beta_{z_n}$ is the multinomial distribution parameter for $z_n$; $\bar{z} = (1/N) \sum_{n=1}^{N} z_n$. We follow the generative process and E-M procedure (Mcauliffe and Blei, 2008) to infer the unknown parameters in the topic and word distributions.

The first comparison is between DNN, ANN, LSTM, and SVMs in which the input features are the selected individual and paired token features. From Figure 7, we can see that the results show that DBN has an overall better performance than the other methods. The performance of SVMs is steady and the results of ANN and SVMs are getting close to that of DBN when $\phi$ is higher than 0.15.

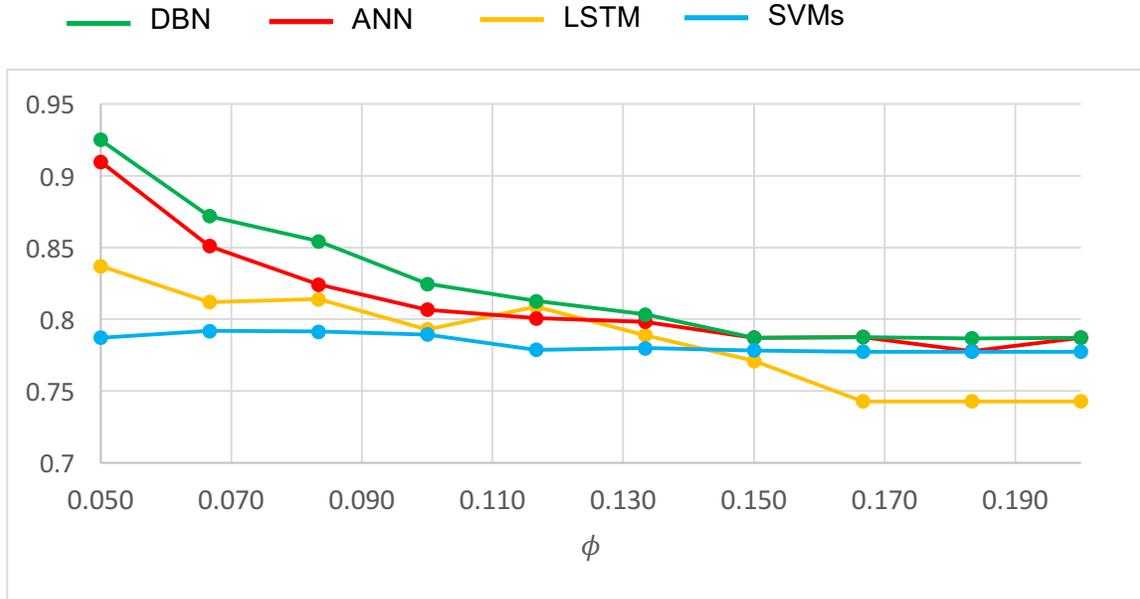

Figure 7 Accuracy comparison between DBN, ANN, LSTM and SVMs with selected individual and paired tokens under different thresholds of correlation coefficient $\phi$.

The second comparison is made between deep learning methods, ANN, SVMs and sLDA when we set $\phi$ as 0.083. From Figure 8, one can see the great advantages of deep learning methods in tweet classification is also better than sLDA.

One can see that both sLDA and LSTM do not perform well as we notice in previous works. Compared to the applications like film reviews, image, etc., tweets have fewer words. Thus, the classification may neither generate reliable topic distribution in sLDA nor give enough input features for LSTM. The words (token) in the tweet posts are usually ad-lib and in a random order: words in a tweet are not well organized; under this condition, LSTM may not be accurate because it depends on the sequential information input at each step. It is also worth mentioning that in sLDA, the input is the original tweets instead of selected features and the results are not affected by $\phi$ values. Thus, the result of sLDA is not put in the comparison in Figure 7.

DBN also gains overall better performance than the SVMs and ANN: compared with DBN, SVMs has an overall lower precision and accuracy while ANN gains relatively lower accident-related precision and higher non-accident-related precision but its overall performance is slightly left behind by DBN in accuracy. This may be due to that from Section 5.1, DBN has more layers and neurons than SVMs and ANN: After several epochs of fine-tuning, DBN can achieve the best combinations of functions and parameters to gain a better result. One can also see that advantages of DBN over SVMs and ANN becomes larger when there are more token features in the model. This is because Deep Learning method is designed to deal with problems such as image processing and speech recognition when there is a large quantity of features and tweets can inherently generate a large number of token features. In this way, DBN can be a better method to process the tweet data than SVMs and ANN.

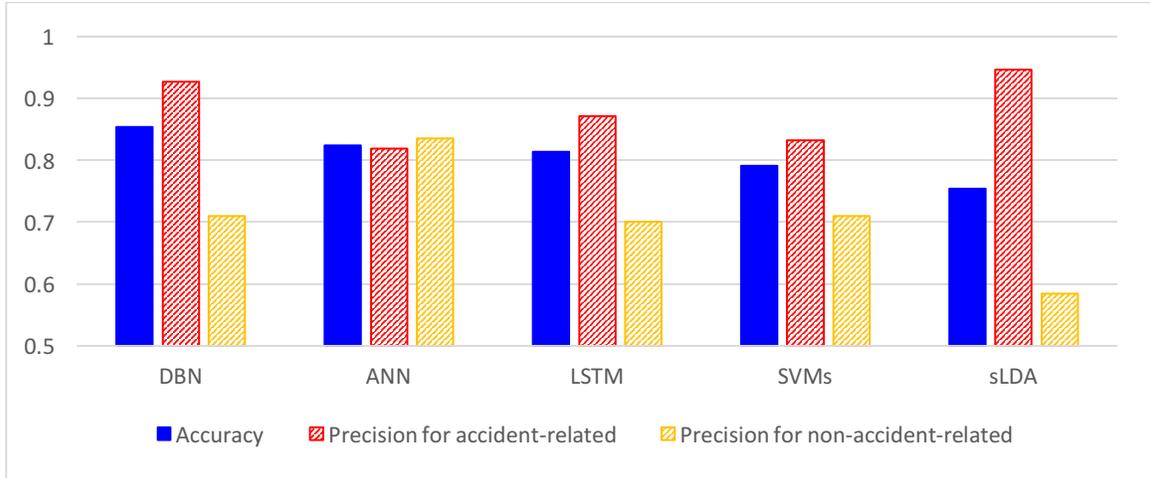

Figure 8 Comparison of accuracy and precision among DBN, ANN, LSTM, SVMs and sLDA when $\phi$ is equal to 0.083.

## 6 Validation with Accident Log and Traffic Data

### 6.1 Coverage of accident-related tweets

In this study, we only consider the geo-tagged tweets, which is an only small portion of entire tweet message population. Given the low coverage of geo-tagged tweets and the disadvantages of non-geo-tagged tweets, we need to admit that tweets are unlikely to cover all the traffic accidents with high probability. Thus, the tweets are more probable to be a viable supplement rather than a replacement to the existing detection method. This is mainly because they are relatively small-scale incidents (Schulz et al., 2013) and seldom arouse public attention. The influence of them may not be as high as that of earthquake or festival parades. Also, not all travelers are willing to leave a corresponding message online. When passing by the site of a traffic accident, most of the drivers cannot tweet for their own traffic safety consideration.

### 6.2 Validation by the traffic accident log on freeway

Given the unpredictability and complexity of social networks, it is still possible that some accident-related tweets are false alarms. Thus, it would be a worthwhile topic to test the credibility of these tweets as well as the time and location effectiveness.

The first comparison is made between labeled tweets and traffic accident log filed by Virginia Department of Transportation (VDOT). The accident log only covers the accidents occurred on the freeway on January, February, March, April, June, July, August and October in 2014. Each accident record in the log has an accident type, a location information (latitude, longitude), start time and end time of the accident. In the comparison, there are 4 major freeways: "I-395", "I-495", "I-66", "I-95" and 3 types of accident: "collision", "disabled vehicle" and "vehicle on fire" (same as what we mentioned in Section 3.2).

It is worth mentioning that using only the geographic latitude and longitude to map-match the locations into road networks or even points of interest (accident log locations) may not be accurate. Caused by GPS errors, the inaccuracy of map-matching is an everlasting problem. Even though, this step may still give insights to the accident studies based on Twitter data. Our results show that there are 110 accident-related tweets map-matched into the freeways and there are 73 (66%) map-matched accident-related tweets by setting the maximum distance difference as 4 miles and the maximum time difference as 1 hour. Setting the time and distance difference is not mandatory but comes through several considerations: given the vast road networks in NOVA, this gap in distance on freeway should be larger than that by (Gu et al., 2016); also, according to the accident start and end in the log, one can see that some accidents on the freeway need 1 or 2 hours to clear out. For the map-matched accident-related tweets: the average distance gap between tweet and accident log is 1.8 miles; also, given that the accident start time is when the police arrives, one can see that there are 23 tweets (31%) posted before the accident starts, 16 (21.9%) during the accident, and 34 (46.6%) after the accident end. Table 7 lists 6 cases of the map-matched tweets.

Table 7 lists 6 tweets map-matched by traffic accident log

| Tweet* | Accident duration | Tweet time versus accident start time |
|---|---|---|
| is the hanson lane at braddock incident resolved | 42 m | 50 m after |
| tailgating with the party bus before the dave matthews band concert | 31 m | 34 m after |
| used to seeing accidents but not everyday you see a 5 car bumper to bumper collision lol | 39 m | 11 m before |
| i just got rear ended on 95 and my truck ate that shit theres not even a scratch thank god for my hitch | 1 h 37 m | 16 m after |
| wtoptraffic accident at w267 between 495 amp route 7 toll plaza all lanes blocked | 16 m | 2 m after |
| crazy car flip accident this morning on commerce st | 1 h 19 m | 1 h 3 m after |

* some special characters or icons have been removed from the tweets.

Besides, there are also 37 accident-related tweets that cannot be map-matched by the accident log and part of them are listed in Table 8. One can see that most of them give a clear expression of accidents but their locations and tweeting times do not match any record in the traffic accident log.

Table 8 lists 6 tweets not map-matched by traffic accident log

| Tweet |
|---|
| stuck on 95 cause an accident arrrggghhh freakin i 95 s |
| 66 is a complete mess this morning feds everywhere couples accidents too |
| accident inner loop just past braddock rd express lane entrance left lane blocked vatraffic |
| graphic photo horrible tractor trailer accident with two dead closes nb 95 in dumfries vatraffic breakingnews |
| wtoptraffic 66e before washington blvd multiple car accident in left lane |
| wow just witnessed a drunk driving accident |

We further examine the tweets including users, tweeting time and locations and brought up several important issues from both Table 7 and 8:

First, one can see the names of influential users or hashtags (vatraffic, wtoptraffic, etc.) from both Table 7 and 8. Our examinations show that out of 73 map-matched and 37 non-map-matched tweets, there are respectively 16 and 10 of them that contain the names of influential users or hashtags. We manually checked the tweets with these names and found that some of them are posted by the reporters working for media or authorities while others are just normal tweet users.

The locations from reporters can sometimes be valid as some of them are airborne reporters or just on the accident sites. Thus, we can conclude that effective accident locations on the freeway may also be acquired from tweets with the authority names.

Second, by checking the location differences, one can see that there are also a number of tweets without authority or hashtag names that cannot be map-matched by accident log. The latitude and longitude provided by tweets may not necessarily be the accident site: this problem on the freeway may be due to that tweets are posted after driving 1 or 2 miles after they saw an accident. Also, the GPS errors and problems of map-matching may also enlarge the distance gaps between the tweets and accident sites. One can also see some location information (95, Dumfries, etc.) posted by the tweets too. However, under most conditions, the tweet location information may not be complete enough relate to an exact location on the freeway.

Third, the time differences between the tweets and accident log are also worth discussing. One can see from Table 7 that some accidents last longer than 1 hour. Some of the tweets are posted even earlier than the accident start time. The posting time of social media in detecting traffic accidents may be a viable topic in future research.

Finally, the possibility of false alarms should not be ignored given a large number of geo-tagged tweets and unpredictable circumstances in the vast road networks in NOVA. Generally, the false alarms, as well as the location errors and map-matching problems complicate the applications of event detection from social media

**6.3 Validation by traffic loop-detector data**

From Section 6.2, one of the major problems is the location difference between the tweets and accident sites. The traffic accident effects may be transmitted to the downstream links and even impact the surrounding networks. Thus, the surrounding traffic where the tweets are posted may be disturbed.

Thus, the second comparison is made between the accident-related tweets and loop-detector data in NOVA. This method referred to a detection method as introduced in Zhang et al. (2016c) which based on the fundamental diagram; it leverages the historical traffic flow and occupancy over the whole year of 2014 recorded by over 15000 detectors in more than 1250 signalized intersections in NOVA and two abnormal indexes are derived to quantify the changes in traffic as compared to its historical records.

In the first step, the traffic occupancy in each detector into *N* separate groups. The medians of the historical traffic flow values over the same occupancy group are defined as the traffic signature of the detector: $F^d = (f_1^d, f_2^d, \cdots, f_n^d)$ where $f_o^d$ is the median value of traffic flow given a range of occupancy $o$ in detector $d$. the K-means algorithm (Münz et al., 2007) and Akaike information criterion (AIC) (Akaike, 1998) are employed to cluster the traffic signatures. AIC value will theoretically decrease with the increase of number of clusters and we obtain the best numbers of clusters as 15 when the change in AIC goes lower than 3%. Figure 9(a) shows the clear differences between the clusters while Figure 9(b) shows the distributed features of traffic flow over the occupancy in on cluster.

In the second step, the deviation degree of the traffic flows to their cluster center is quantified over an occupancy interval in each cluster. Previous empirical examinations show that the distributions of the traffic flows over an occupancy interval follow a Gaussian distribution (Zhang et al., 2016a) which can be formulated as Equation.

$$P^{dt} = \Phi\left(\left|\frac{\mathcal{F}_o^{dt} - C_o^i}{\sigma_o^i}\right|\right) \tag{15}$$

Where $P^{dt}$ is the abnormal degree which is the probability for detector $d$ over time period $t$. $i$ indicates the $i$th cluster of $d$; $\mathcal{F}_o^{dt}$ is the traffic flow data over traffic occupancy interval $o$; $\sigma_o^i$ and $C_o^i$ is the standard deviation and center of traffic flow in Cluster $i$ over occupancy interval $o$.

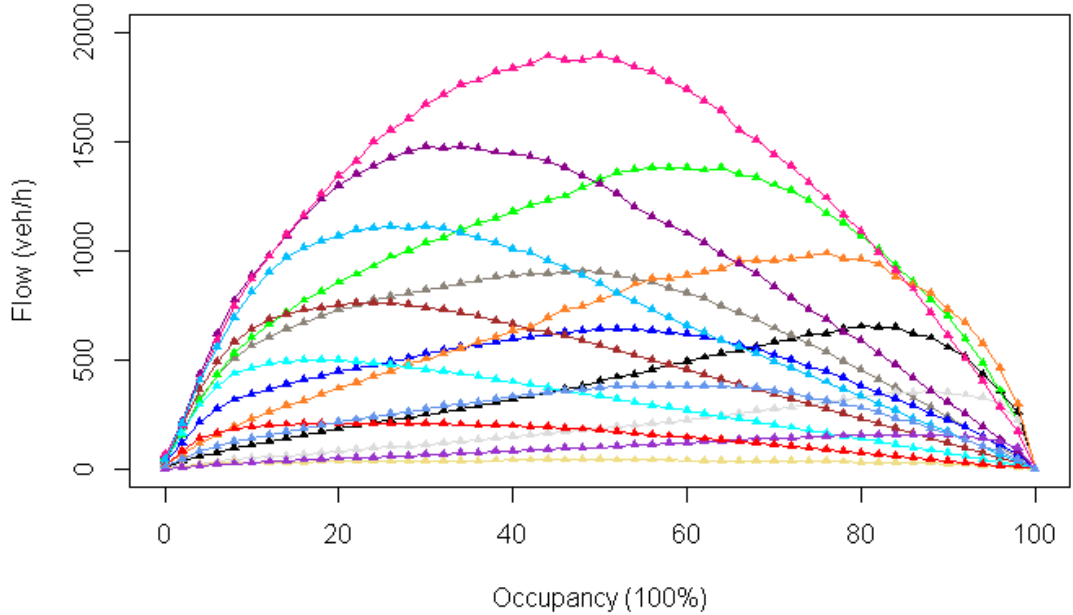

(a)

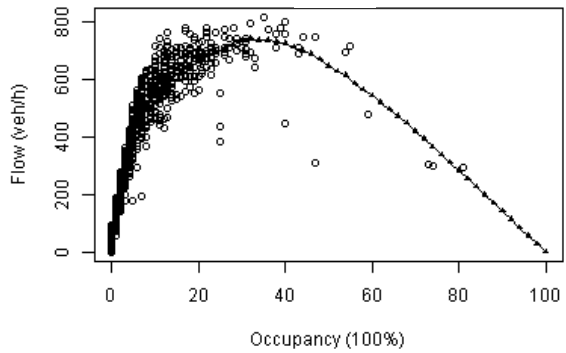

(b)

Figure 9 (a) 15 different cluster centers of traffic signatures. (b) Comparison between cluster centers and the original traffic flow and occupancy data in one sample detector

In the third step, we pair the tweets in the local roads with the abnormal degrees of nearby detectors. The temporal ranges are set to be before and after 1 hour when the tweet is blogged and the spatial range is set to be 1 mile. By this setting, there are 231 qualified tweets that can be paired with the abnormal degrees. For each tweet, we aggregate the abnormal degrees in its nearby detector in two ways: mean value and 75th quartile value:

$$p_{traffic} = \frac{1}{NUM} \sum_{t \in dom(t)} \sum_{d \in dom(d)} P^{dt} \qquad (16)$$

$$q_{traffic} = Q3(\{P^{dt}, d \in dom(d) \cap t \in dom(t)\}) \qquad (17)$$

Where $t$ is the hour period; $d$ is the detector ID and $i$ is the cluster ID; $dom(d)$ is the domain of all the detectors within the geo-scale of the tweets and $dom(j)$ is the domain of all time periods within the time-scale of the tweets; $Q3()$ is the operator of 75th percentile; $NUM$ is the total number of traffic observations related to a tweet.

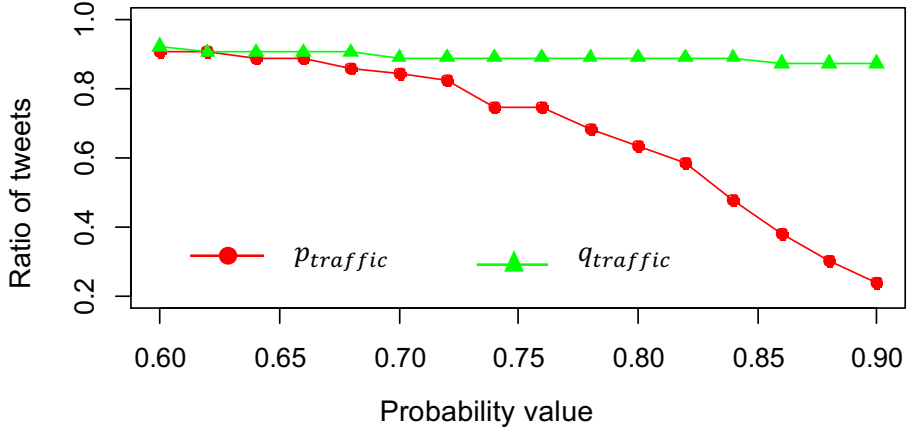

(a)

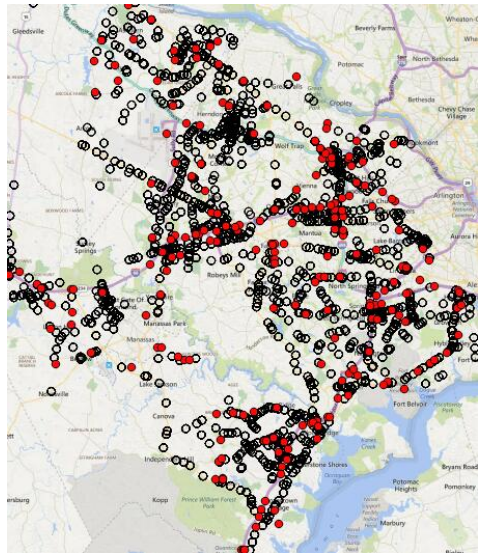

(b)

Figure 10 (a) Ratio of tweets whose surrounding detectors have $p_{traffic}$ and $q_{traffic}$ higher than a probability value (b) geographic distribution comparison of locations between tweets and loop detectors. For these tweets, one can find loop detectors within 1 mile.

Figure 10(a) shows the ratio of tweets that can find $p_{traffic}$ and $q_{traffic}$ higher than a certain value while Figure 10(b) shows the geography of detectors and accident-related tweets. According to Equation 16, probability equal or near 0.5 should be abnormal. The higher $p_{traffic}$ or $q_{traffic}$ is, the worse the traffic condition should be. One can see that for most accident-related tweets, the ratio drops dramatically when $p_{traffic}$ increases. This can be interpreted as the overall surrounding traffic can be slightly affected in most accidents and severely affected by some of them. Also, the ratio decreases slightly when $q_{traffic}$ is getting smaller, this means that there are some detectors around the tweet locations record traffic conditions while others do not. In conclusion, if the probability is set to be 0.9, there are more 80% of the tweets that can be traced by the abnormal traffic.

Recalling the comparison between tweets and accident log, it is entirely possible that the tweets can capture the unexpected small events happened in our daily life. These events may include those "mild" accidents that do not incur the attention of traffic police and thus may not be included in the official log. The consequences of these events such as the road lanes blocking or cars slowing down may not last long, and the corresponding affairs may come with a proper handling. If so, the untraceable tweets may act as a secondary tool to the current accident detection system.

**7 Conclusions and discussions**

In this paper, we employ deep learning in detecting traffic accidents from social media data. Our intentions are detailed by three major steps: Feature Selection, Classification, and Validation. The classification results show the great advantages of Deep Belief Network (DBN) over LSTM, ANN, SVMs, and sLDA. The validation via accident log and loop-detector data shows some unique time and space features of social media. Our findings can be summarized as follows:

First, we thoroughly investigate the tweet contents related to traffic accidents. We found token features: individual tokens and paired tokens that may indicate the event of a traffic accident. Our results show that paired tokens can capture the association rules inherent in the accident-related tweets. The paired token features can further increase the accuracy of the traffic accident detection, especially when number of available individual token features is limited.

Second, DBN can obtain an overall accuracy of 85% with 44 individual token features and 17 paired token features. Further, DBN outperforms the ANN with one hidden layer, the sequence labeling with LSTM, as well as the traditional method SVMs. The gap is even bigger when there are more token features. It can also obtain a better accuracy than the popular topic modeling method sLDA. The results verify that deep learning has advantages in classifying disorderly short texts in tweets.

Finally, comparisons between the accident-related tweets with both the traffic accident log and loop-detector data indicate some merits of tweets. It is found that on freeways, nearly 66% of the accident-related tweets on freeways can be located by the accident log. The effectiveness of time and location of Twitter, as well as the functionality of hashtags and influential users have been fully discussed for accident detection using Twitter. Besides, on local roads, more than 80% of them can be related to the surrounding abnormal traffic data. This may indicate some of the tweets capture some accidents that are not documented by the police and are worth studying in the future.

In sum, integrating social media data into the traffic-related study opens up a wide range of possibilities for transportation research. The results show that social media data might be noisy and even unreliable. Therefore, social media in accident detection can function as a secondary source rather than a replacement to the traditional method. The potential of finding "unrecorded" accidents shows the power of massive wisdom collection given by social media. The model calibrated in this study can be employed to detect the traffic accident in a real-time manner, which potentially lead to better emergency responses. Even more accurate models can be calibrated in the future by a joint community effort in creating a dataset so that it could be commonly used for research, following the example of the domain of information retrieval (Kuflik et al., 2017). Further studies can focus on the data fusion of different data sources to better realize the purposes of other research such as traffic jam detection, traffic control in emergency evacuation (Asamoah, 2014), etc. The spatial-temporal features of traffic data are also worth studying for regional traffic operations (Zhang et al., 2016b). Note that our tweet data are geotagged. It would also be an interesting extension to detect traffic events with non-geotagged tweets such as those posted by 511 Traffic Systems (DMV.org, 2017).

**Acknowledgement**

This research was supported by Transportation Informatics (TransInfo) University Transportation Center at University at Buffalo, The State University of New York. The traffic data and accident log were provided by Jizhan Gou and Xiaoling Li from VDOT. Authors appreciate their data support.

Cramér, H., 1999. *Mathematical methods of statistics*. Princeton university press.

D'Andrea, E., Ducange, P., Lazzerini, B., Marcelloni, F., 2015. Real-time detection of traffic from twitter stream analysis. *Intelligent Transportation Systems, IEEE Transactions on* 16(4), 2269-2283.

Deng, L., Yu, D., 2014. Deep learning: methods and applications. *Foundations and Trends® in Signal Processing* 7(3–4), 197-387.

DMV.org, 2017. 511 Traffic Systems.

Freund, Y., Schapire, R.E., 1999. Large margin classification using the perceptron algorithm. *Machine learning* 37(3), 277-296.

Gal-Tzur, A., Grant-Muller, S.M., Kuflik, T., Minkov, E., Nocera, S., Shoor, I., 2014. The potential of social media in delivering transport policy goals. *Transport Policy* 32, 115-123.

Geisser, S., 1993. *Predictive inference*. CRC Press.

Graves, A., 2012. Supervised sequence labelling, *Supervised Sequence Labelling with Recurrent Neural Networks*. Springer, pp. 5-13.

Graves, A., Liwicki, M., Fernández, S., Bertolami, R., Bunke, H., Schmidhuber, J., 2009. A novel connectionist system for unconstrained handwriting recognition. *IEEE transactions on pattern analysis and machine intelligence* 31(5), 855-868.

Graves, A., Mohamed, A.-r., Hinton, G., 2013. Speech recognition with deep recurrent neural networks, *Acoustics, speech and signal processing (icassp), 2013 ieee international conference on*. IEEE, pp. 6645-6649.

Greff, K., Srivastava, R.K., Koutník, J., Steunebrink, B.R., Schmidhuber, J., 2016. LSTM: A search space odyssey. *IEEE transactions on neural networks and learning systems*.

Gu, Y., Qian, Z.S., Chen, F., 2016. From Twitter to detector: Real-time traffic incident detection using social media data. *Transportation Research Part C: Emerging Technologies* 67, 321-342.

Hahsler, M., Grün, B., Hornik, K., 2007. Introduction to arules–mining association rules and frequent item sets. *SIGKDD Explor*.

Hall, F.L., Shi, Y., Atala, G., 1993. *On-line testing of the McMaster incident detection algorithm under recurrent congestion*.

Hasan, S., Ukkusuri, S.V., 2014. Urban activity pattern classification using topic models from online geo-location data. *Transportation Research Part C: Emerging Technologies* 44, 363-381.

He, Q., Kamarianakis, Y., Jintanakul, K., Wynter, L., 2013. Incident duration prediction with hybrid tree-based quantile regression, *Advances in Dynamic Network Modeling in Complex Transportation Systems*. Springer, pp. 287-305.

Hinton, G.E., 2009. Deep belief networks. *Scholarpedia* 4(5), 5947.

Hinton, G.E., Osindero, S., Teh, Y.-W., 2006. A fast learning algorithm for deep belief nets. *Neural computation* 18(7), 1527-1554.

Hochreiter, S., Schmidhuber, J., 1997. Long short-term memory. *Neural computation* 9(8), 1735-1780.
27